# Imaging Magnetization Structure and Dynamics in Ultrathin YIG/Pt Bilayers with High Sensitivity Using the Time-Resolved Longitudinal Spin Seebeck Effect

Jason M. Bartell[1], Colin L. Jermain[1], Sriharsha V. Aradhya[1], Jack T. Brangham[2], Fengyuan Yang[2], Daniel C. Ralph[1,3], Gregory D. Fuchs[1]

[1]Cornell University, Ithaca, NY 14853, USA

[2]Department of Physics, The Ohio State University, Columbus, OH 43016, USA

[3]Kavli Institute at Cornell for Nanoscale Science, Ithaca, NY 14853, USA

Abstract

We demonstrate an instrument for time-resolved magnetic imaging that is highly sensitive to the in-plane magnetization state and dynamics of thin-film bilayers of yttrium iron garnet ($Y_3Fe_5O_{12}$, YIG)/Pt: the time-resolved longitudinal spin Seebeck (TRLSSE) effect microscope. We detect the local, in-plane magnetic orientation within the YIG by focusing a picosecond laser to generate thermally-driven spin current from the YIG into the Pt by the spin Seebeck effect, and then use the inverse spin Hall effect in the Pt to transduce this spin current to an output voltage. To establish the time resolution of TRLSSE, we show that pulsed optical heating of patterned YIG (20 nm)/Pt(6 nm)/Ru (2 nm) wires generates a magnetization-dependent voltage pulse of less than 100 ps. We demonstrate TRLSSE microscopy to image both static magnetic structure and gigahertz-frequency magnetic resonance dynamics with sub-micron spatial resolution and a sensitivity to magnetic orientation below 0.3 deg/$\sqrt{Hz}$ in ultrathin YIG.

Main Text

Ultrathin bilayers of the magnetic insulator YIG interfaced with a heavy, non-magnetic metal (NM) such at Pt are being intensely studied for the development of high-efficiency magnetic memory and logic devices operated by spin-orbit torque [1,2], for magnon generation and propagation [3–5], and as a model system for understanding spin-current generation by the longitudinal spin Seebeck effect (LSSE) and spin pumping [6–9]. For all of these research areas, it would be useful to have a high-sensitivity and local probe of magnetization dynamics in the YIG layer, especially for the ultrathin films required in many devices. This has proven challenging, and although magneto-optical techniques such as Brillouin light scattering and the magneto-optical Kerr effect (MOKE) have proven valuable [3,10–14], they have not enabled direct time-resolved imaging of magnetic precession or direct imaging of in-plane magnetization of ultra-thin YIG films (20 nm and below). An alternative approach that enables in-plane imaging of YIG/Pt bilayer devices was demonstrated by Weiler *et al*. [15]. In that work, the authors use laser heating to image the in-plane magnetic structure of YIG, but not its dynamics. Here we extend the approach into the time domain to perform high sensitivity imaging of the in-plane magnetic orientation ($< 0.3°/\sqrt{Hz}$) with sub-micron spatial resolution and sub-100 ps temporal resolution. Using TRLSSE microscopy we can observe, for example, that the resonance field in ultra-thin YIG films can vary by up to 30 Oe within micron-scale regions of a YIG/Pt device. Our results demonstrate that TRLSSE microscopy is a powerful tool to characterize static and dynamic magnetic properties in ultrathin YIG.

The principle behind the TRLSSE microscope, shown schematically in Fig. 1, is the generation and detection of a thermally generated local spin current [16]. For the case of YIG/Pt, a local thermal gradient perpendicular to the film plane is generated by laser heating of Pt. The gradient creates a thermally-induced spin current that is proportional to the local magnetization [17–19]. The spin current that flows into the Pt is detected with the ISHE [20,21] in which spin-orbit coupling leads to a spin-dependent transverse electric field. For this work, the resulting voltage can be described as [17,19] $V_{LSSE} \propto -\xi_{SH} S \frac{\mathbf{M}(x,t)}{M_s} \times \nabla \mathbf{T}(x,t)$, where, $\xi_{SH}$ is the spin Hall efficiency, $S$ is the spin-Seebeck coefficient, $\mathbf{M}$ is the local magnetization, $M_s$ is the saturation magnetization and $\nabla \mathbf{T}$ is the thermal gradient. The LSSE has been attributed to both thermal gradients across the thickness of YIG and to interfacial temperature differences between YIG and Pt [17–19,22,23]. Our experiment cannot definitively distinguish between these two mechanisms. Thus, here we discuss only $\nabla \mathbf{T}$ as single quantity for simplicity and for consistency with our prior work using the anomalous Nernst effect, however, this question requires further study. $V_{LSSE}$ is a read-out of the local magnetization $m_y$ because the electric field is generated in response to the spatially local z-component of the thermal gradient, $\nabla T_z$ (coordinates as defined in Fig. 1) [15,24].

To extend LSSE imaging into the time-domain, we use picosecond laser heating to stroboscopically sample magnetization. We have previously shown, in metallic ferromagnets, that picosecond heating can be used for stroboscopic magnetic microscopy using the time-resolved anomalous Nernst effect (TRANE) [25]. In TRANE microscopy, the temporal resolution is set by the excitation and decay of a thermal gradient within a single material that both absorbs the heat from the laser pulse and produces a TRANE voltage from internal spin-orbit interactions [26,27]. In the LSSE however, the timescale of spin current generation can depend on both the timescale of the thermal gradient and the timescale of energy transfer between the phonons and magnons. Recent experiments indicate that in the qausi-static regime the magnon-phonon relaxation rate may play a dominant role [28–31]. Using picosecond heating and time-resolved electrical detection to move beyond the quasi-static regime, we show a TRLSSE in agreement with a recent all-optical experiment [22].

We grew our samples using off-axis sputtering onto (110)-oriented gadollinum gallium garnet ($Gd_3Ga_5O_{12}$, GGG), [32–34] followed by *ex situ.* deposition of 6 nm of Pt with a 2 nm Ru capping layer. Photolithography and ion milling were used to pattern wires and contacts for wirebonding. We present measurements of a 2 µm × 10 µm wire and a 4 µm × 10 µm wire with DC resistances of 296 Ω and 111 Ω respectively. In this room temperature study, we neglect the potential anomalous Nernst effect of interfacial Pt with induced magnetization [35,36], and we neglect a possible photo-spin voltaic effect [37], neither of which can be distinguished from TRLSSE in presented measurements.

Our TRLSSE measurement consists of pulsed laser heating and homodyne electrical detection as shown in Fig. 2a. We use a Ti:Sapphire laser pulse to locally heat the sample with 3 ps pulses of 780 nm light at a repetition rate of 25.5 MHz. The electrical signal produced at the sample is the sum of the LSSE dependent voltage, $V_{LSSE}(\nabla T_z, \mathbf{M})$, and a voltage, $V_J(\Delta T, J)$, which is generated when a current density $J$ is passing through the local region of Pt with increased resistance due to laser heating [38]. To reject noise and recover the signal of the resulting electrical pulses, we use a time-domain homodyne technique in which we mix the $V_{LSSE} + V_J$ pulse train with a synchronized reference pulse train, $V_{mix}$, in a broadband (0.1-12 GHz) electrical mixer. The mixer output is the convolution of the two pulse trains given by [38]

$$V_{sig}(\boldsymbol{x}, \tau) = K \int_0^\Gamma (V_{LSSE}(\nabla T_z(\boldsymbol{x}, t), \mathbf{M}(\boldsymbol{x}, t)) + V_J(\Delta T(\boldsymbol{x}, t), J(\boldsymbol{x}, t)) V_{mix}(\tau - t) dt, \qquad (1)$$

where $x(x,y)$ is the laser spot position in the sample plane, $\Gamma$ is the period of the laser pulses, $K$ is the transfer coefficient, and $\tau$ is the relative delay. A relative delay of zero corresponds to the maximum of both pulse trains arriving at the mixer simultaneously.

We study the timescale of the LSSE signal generated by a picosecond pulse by measuring $V_{sig}$ as a function of mixer delay $\tau$. Fig. 2b shows the result of this measurement using a 100 ps mixing pulse reference, $V_{mix}$, at a saturating magnetic field, $H$, perpendicular to the wire at $H = +414$ Oe and $-414$ Oe, respectively. In Figure 2c we plot the difference between these two voltage traces to reject non-magnetic contributions. We find that the full-width at half-maximum (FWHM) is $100\pm10$ ps, which is followed by electrical oscillations that we attribute to non-idealities in the detection circuit (see the SI for further discussion.) Because the duration of the magnetic component of $V_{sig}$ is experimentally indistinguishable from the FWHM of $V_{mix}$, we conclude that 100 ps is an experimental upper bound for the TRLSSE signal duration. To our knowledge, this is the first direct electrical measurement of picosecond duration LSSE voltages.

To calibrate the local change in the Pt temperature, $\Delta T_{Pt}$, due to picosecond heating and to quantify the rate of thermal relaxation, we measure $V_J$ in the presence of a DC current, which uses the local Pt resistivity as an ultra-fast thermometer. Figure 2d shows $V_J$ as a function of mixer delay, $V_J(\tau) = V_{sig}(\tau, J = 4.2 \text{ MA/cm}^2) - V_{sig}(\tau, J = -4.2 \text{ MA/cm}^2)$, for applied currents of $\pm 0.5$ mA. $V_J(\tau)$ is proportional to $\Delta T_{pt}$ through $V_J$, but it is not proportional to either the magnetic state of the sample or $\nabla T_z$. We observe that $V_J$ relaxes to zero faster than the laser repetition period, indicating that the sample thermally recovers between pulses. To quantitatively consider the spatiotemporal thermal evolution, we performed a time-domain finite element (TDFE) calculation of focused laser heating in the wire. Additional details are available in the SI, and see Ref. [25] for a lengthier discussion of the procedure. The comparison of the spatiotemporal profile of the calculation and the known temperature dependence of resistivity enable us to calibrate the spatiotemporal temperature rise due to laser heating. We find that the peak film temperature changes by ~50 K in the platinum and ~ 10 K in the YIG for a laser fluence of 5.8 mJ/cm$^2$, which is the maximum for the presented measurements. Note that we assume all laser heating is mediated by optical absorption in Pt because YIG and GGG are transparent at 780 nm [39,40]. The TDFE calculation reveals that, in agreement with experiment, $\nabla T_z$ across the YIG thickness decays more quickly than the full thermal relaxation of the Pt back to the ambient temperature (e.g. $\Delta T_{pt} = 0$). This difference in timescales between $\nabla T_z$ and $\Delta T_{pt}$ is important because the magnetic signal in our experiment is sensitive to only $\nabla T_z(t)$, not $\Delta T_{pt}(t)$ of the Pt.

The sub-100 ps spin current lifetime in our experiment is short enough that the TRLSSE is useful for stroboscopic measurements of resonant YIG magnetization dynamics. To confirm this idea, we use TRLSSE microscopy to measure ferromagnetic resonance (FMR) by driving a gigahertz-frequency a.c. current into the Pt, which generates magnetic torques on YIG from both the Oersted magnetic field and from spin currents generated by the spin Hall effect [41–43]. The current is generated with an arbitrary waveform generator (AWG) that is phase-locked to the laser repetition rate and coupled to the YIG/Pt device through a circulator (see schematic in Fig. 3a). Synchronizing the a.c. current and the laser repetition rate ensures a constant but controllable phase between the precessing magnetization and the sensing heat pulse for a given driving frequency and magnetic field. In our FMR measurements, we fix $\tau = 0$ and align the wire axis parallel to the external magnetic field. In this configuration, the TRLSSE signal is stroboscopically sensitive to the magnetic projection $m_y$ at a particular phase of the magnetic precession about the *x*-axis. In addition to $V_{LSSE}$, $V_{sig}$ contains a contribution from $V_J$ that is proportional to the local a.c. current amplitude and phase [38]. We separate the magnetic $V_{LSSE}$ from the non-magnetic $V_J$ by measuring $V_{sig}$ with a lock-in amplifier referenced to a 383 Hz, 7.6 Oe RMS modulation of the external magnetic field. Fig. 3b shows LSSE FMR spectra as a function of field that is excited using a 0.5 mA a.c. current at 4.1 and 4.9 GHz. In the limit that the modulation magnetic field is small compared to the FMR linewidth, we can interpret the resulting signal $V_{mod}$ as a derivative signal that contains a linear combination of the real and imaginary parts of the dynamic susceptibility, $\chi$, $V_{mod}(H) \propto \frac{d\chi'}{dH} Sin(\theta) + \frac{d\chi''}{dH} Cos(\theta)$. This relation is used to fit the FMR spectra to extract the amplitude, phase, linewidth, and resonant field. For more details on fitting see refs [25,38]. To demonstrate that the TRLSSE microscope is a phase-sensitive stroboscope, we rotated the phase of the microwave current by 180° and re-measure FMR. As expected, inverting the phase of the drive inverts the phase of the FMR lineshape (Fig. 3c).

Next, we quantify the sensitivity of TRLSSE microscopy for our ultra-thin YIG/Pt samples. Figure 4 shows representative LSSE measurements of the YIG magnetization versus magnetic field perpendicular to the wire at several optical powers. In this geometry, the positive and negative saturation values of $V_{LSSE}$ quantify the full range of magnetization, $+M$ to $-M$. Then, using the standard deviation of the noise in the LSSE voltage, $\sigma_{LSSE}$, we can quantify the angular sensitivity noise floor assuming small angle magnetic deviations from the wire axis, such as for stroboscopic FMR measurements. The sensitivity is calculated using [25] $\theta_{min} = \frac{\sigma_{LSSE}}{\sin(\theta_o)(V_{LSSE}^{max}-V_{LSSE}^{min})/2}\sqrt{TC}$ where $TC$ is the lock-in time constant. We find a sensitivity of 0.3 deg/√Hz for an optical power of 0.6 mW, corresponding to a laser fluence of 5.8 mJ/cm$^2$. It is important to note that the sensitivity is sample dependent through both sample geometry and the impedance match with the detection circuit [25].

The interface quality of the sample plays a key role in determining the sensitivity. As spin current diffuses into the platinum, it is subject to loss at the interface. A good indication of interfacial spin transparency is the spin Hall magnetoresistance (SMR) [44,45], which is sensitive to the spin mixing conductance at the interface. For the data presented here, the devices show a SMR of 0.063%, which is the largest value by a factor of 2 from the other devices we patterned. This is consistent with a number of recent SMR reports [44–48], and we expect the high SMR value indicates strong spin transparency at the YIG/Pt interface. We also studied YIG/Pt samples with no measureable SMR which we expect to have a significantly reduced LSSE induced ISHE voltage. We found that the LSSE signal in these devices is approximately an order of magnitude lower for the same laser fluence. Additional details are in the SI.

Having placed upper bounds on the time resolution and quantified the sensitivity, next we demonstrate the application of TRLSSE microscopy for imaging of static magnetization. We acquire images by scanning the laser focus and making a point-by-point measurement of the TRLSSE voltage and reflected light. Figures 5a and 5b show a reflected light image and saturated LSSE image, respectively, for a 4 µm wide YIG/Pt device. In the reflection image, we see the structure of the wire and the contact pads at both ends. We acquired the TRLSSE image at $H = -405$ Oe and shifted the background level for clarity of the color scale. No other image processing was performed. We observe a uniform magnetization state of the YIG/Pt device, as expected from the previously presented magnetic hysteresis measurements (Fig. 4). When we reduce the field to near zero ($H = 4$ Oe) and re-image the wire (Fig. 5c), magnetic texture is revealed that indicates non-uniform canting of the device magnetization. To more clearly show the variation in contrast between images, we plot line cuts of Figs. 5a-c in Fig. 5d. Despite the inhomogeneous remanence that is evident in Fig. 5c, we were not able to observe domains with oppositely aligned magnetization; possibly because once a reversal domain is nucleated, the domain wall propagates without strong pinning.

Without a 180º domain wall the spatial resolution of TRLSSE cannot be directly evaluated. Nevertheless, we use the reflected light image and TDFE simulations to study the possibility that lateral thermal spreading degrades the resolution. To approximate the lateral point spread function of the laser, we fit a scan of the wire step edge to a Gaussian point spread function. This yields a spot FWHM of 0.606 µm. Calculations of the heating indicate that the thermal gradient does not spread laterally in the Pt, thus we expect that the resolution of the TRLSSE is the same as the diffraction-limited optical resolution in this experiment.

We now demonstrate that TRLSSE microscopy has the sensitivity to image dynamic magnetization in the 4 µm YIG/Pt device, which provides quantitative and spatially localized information about dynamical properties of ultrathin YIG materials. As described above, for FMR characterization we orient the external magnetic field parallel to the wire axis and drive a 1.1 mA, 4.9 GHz current into the wire. We image dynamical magnetization at a series of magnetic fields near the resonance field, from $H$ = 896 Oe to 1105 Oe, and plot a selection of the unprocessed images in Figs. 5e-g. The data show that at $H$ far from resonance (Fig. 5e) where precession amplitudes are tiny, the TRLSSE signal at the center of the wire is well below the detection noise floor. There is a small, current-induced, non-magnetic signal artifact at the edges of the wire which we discuss further in the supplementary information. For $H$ near the resonant field, $H_{res}$, the device has a strong, position-dependent TRLSSE response. To quantitatively analyze the data, images are corrected for background offset and sample drift before fitting a resonance field curve for each pixel. We plot a selection of curves from individual pixels in Fig. 6a. We then construct a spatial map of each fitting parameter: $H_{res}$, relative phase, $\phi$, amplitude, $A$, and linewidth, $\Delta H$, and offset, all of which are shown in Fig. 6b-f. We immediately notice spatial variation in these images that is qualitatively similar to the non-uniform magnetic remanence texture shown in Fig. 5c. Together, these measurements confirm the presence of varying local magnetic anisotropy and quantify both static and dynamic magnetic properties in each region. The ability to quantitatively relate the spatial variation of static and dynamic properties in ultrathin YIG/Pt devices is a unique capability of our microscope.

In conclusion, we have demonstrated sensitive and high-resolution TRLSSE microscopy of ultrathin YIG/Pt devices that we expect will prove useful for developing spintronic applications. Using picosecond heating, we demonstrate that TRLSSE microscopy is a sub-100 picosecond probe of ultra-thin YIG/Pt device magnetization, both for static magnetic configurations and for dynamical measurements at gigahertz frequencies. We have demonstrated an angular sensitivity of $0.3°/\sqrt{Hz}$, which to our knowledge is the most sensitive experimental probe of ultra-thin YIG magnetic orientation reported to date.

## Acknowledgments

We thank J. Kimling and D. G. Cahill for helpful comments on an early version of the manuscript, and for providing the interface thermal resistance of YIG/Pt. This research was supported by the U.S. Air Force Office of Scientific Research, under Contract No. FA9550-14-1-0243, and by U.S. National Science Foundation under Grants No. DMR-1406333 and DMR-1507274 and through the Cornell Center for Materials Research (CCMR) (DMR-1120296). This work made use of the CCMR Shared Facilities and the Cornell NanoScale Facility, a member of the National Nanotechnology Coordinated Infrastructure, which is supported by the NSF (Grant No. ECCS-1542081).

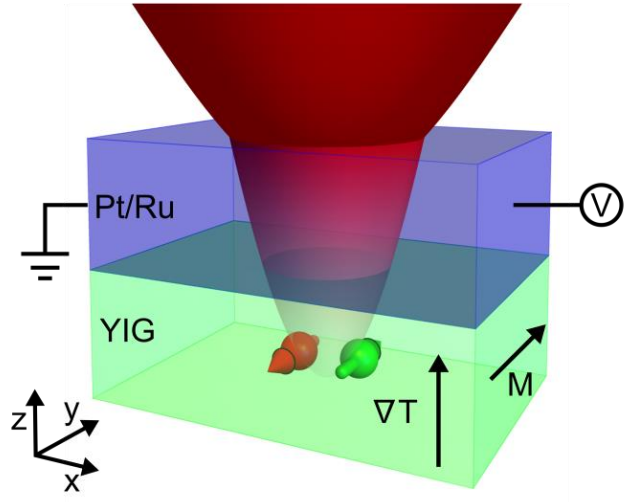

FIG. 1 Schematic of our TRLSSE measurement. A 780 nm, 3 ps pulsed laser, focused to a 0.606 µm diameter spot, is used to heat a YIG (20 nm)/Pt(6 nm)/Ru(2 nm) film. The heating from the laser creates a temperature gradient, $\nabla T_z$. The pulsed heating drives a pulsed magnon flux, $\mathbf{J_s}$, from the YIG into the Pt where it is transduced into a pulsed voltage via the ISHE.

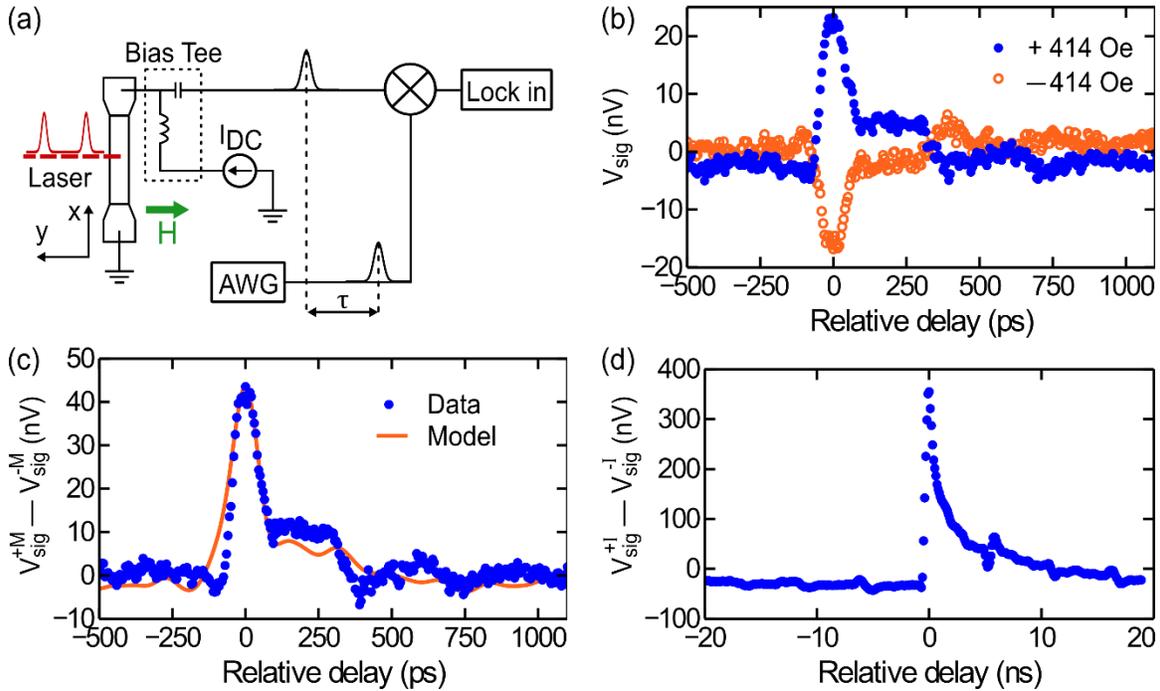

FIG. 2 (a) Schematic of the LSSE detection circuit used for time-resolved voltage measurements. (b) Time-domain measurement of the LSSE generated voltage in the 2 µm wide wire. The time-varying LSSE signal is measured by electrically mixing the pulsed laser generated voltage with a 100 ps voltage pulse from the AWG. Comparing measurements of the YIG at +414 Oe (filled blue circles) and –414 Oe (open orange circles) shows that the signal depends on the orientation of the magnetic moment. Here d.c. level noise and has been removed. The data was acquired with a lock-in time constant of 500 ms and integration time of 2 s per point. (c) The solid blue circles show the difference between the two curves in (b), The orange line is a model, normalized by the data amplitude, of the signal determined by numerically convolving the calculated thermal gradient with the measured mixing pulse. (d) Difference signal of the temperature dependent voltage $V_J$ measured using +/– 0.5 mA and a 600 ps mixing pulse. In (b-d) we report the voltage as detected at the lock-in after passing through the r.f. mixer, not the LSSE signal at the sample itself.

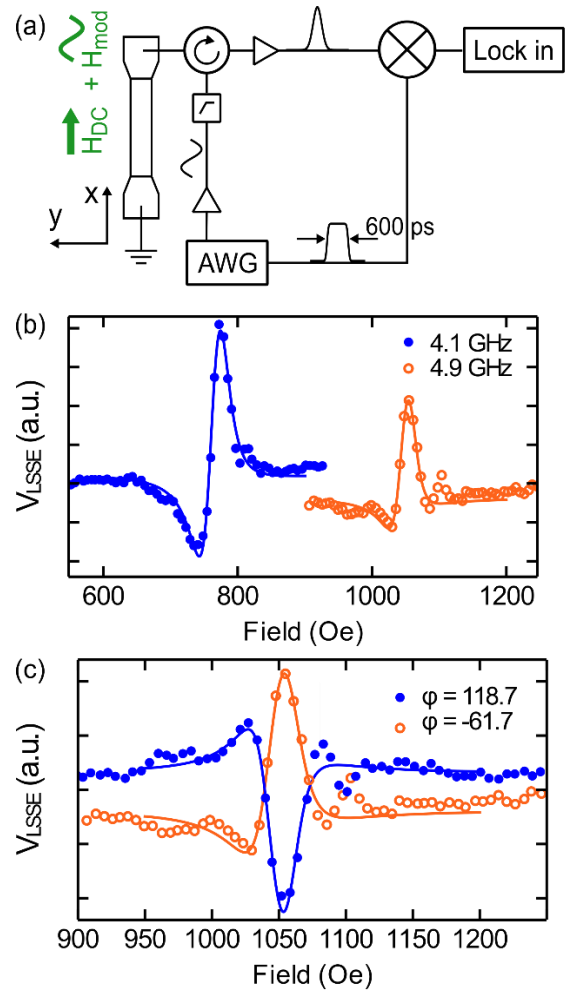

FIG. 3 Stroboscopic detection of ferromagnetic resonance a) Schematic of measurement circuit for detection of magnetization dynamics in the 2 µm wide wire. b) TRLSSE detected FMR for 4.1 GHz (blue, closed circles) and 4.9 GHz (orange, open circles) excitation. The solid lines are a fit to the data using a modified Lorentzian. c) Demonstration of stroboscopic FMR detection in which we measure the response of the YIG driven at phases that differ by 180 degrees. The data was acquired with a lock-in time constant of 1s and integration time of 5 s per point.

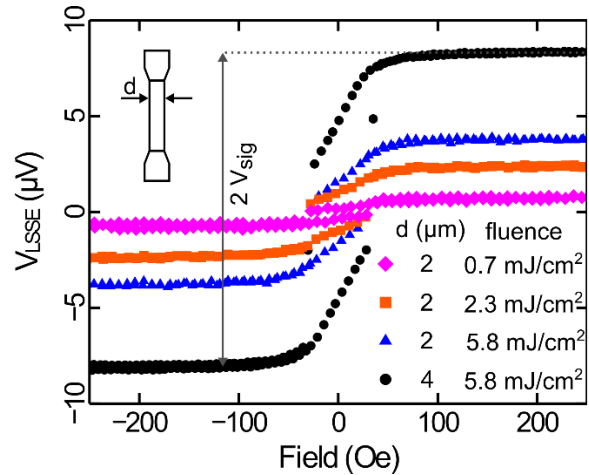

FIG. 4 Measurement of YIG magnetization with LSSE measuring $V_{\text{LSSE}}$ versus external magnetic field for different laser powers and wire widths. For these curves, a DC background was subtracted. The inset shows the wire geometry. We define the signal size to be one-half of the difference in voltage when the magnetization is saturated in opposing directions. The data was acquired with a lock-in time constant of 500 ms and integration time of 2 s per point.

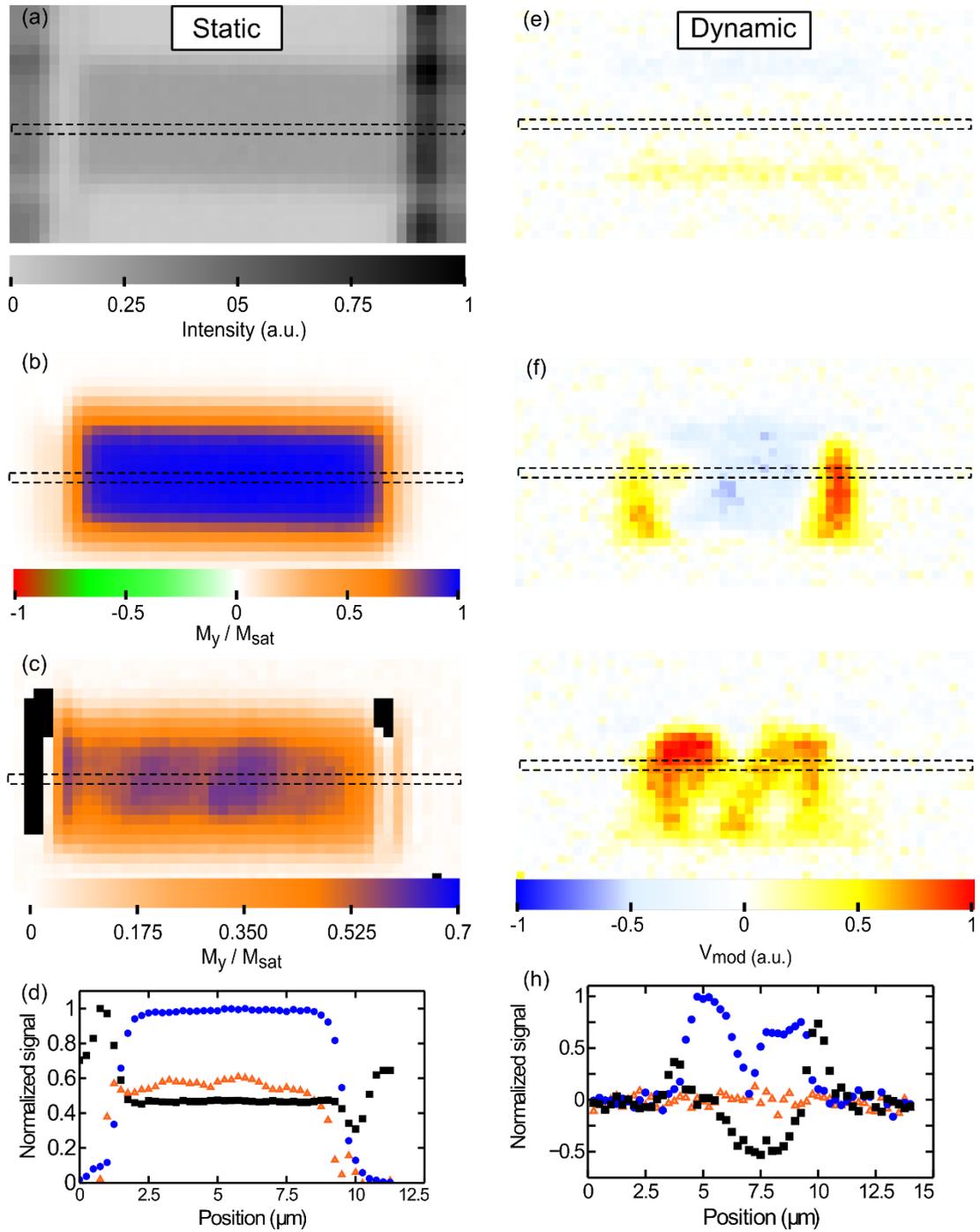

FIG. 5  Images of the 4 µm wide YIG/Pt wire (a) Reflected light image of the YIG/Pt wire measured with a photodiode at the same time as the LSSE voltage. (b) Background subtracted LSSE voltage at saturated magnetization and (c) remnant magnetization at 4 Oe after saturation.

(d) Line cuts of the 2D scans. The normalized reflection signal is shown with black squares, blue circles represent the saturated magnetization, and the orange triangles represent the magnetization of the remnant state. Note, that in the line cuts the low field line cut is normalized with respect to the saturation magnetization. The right side of the figure represents the raw images of the 4 μm wire at different fields around the resonance: (e) 896 Oe. (f) 1007 Oe, (g) 1025 Oe. Images (e-g) share the same color scale. Line cuts of the images are shown in (h) black squares, blue circles, and orange triangles correspond to the boxed regions of (e), (f), and (g) respectively. For (e-g) the data was acquired with a lock-in time constant of 200 ms and an integration time of 2 s.

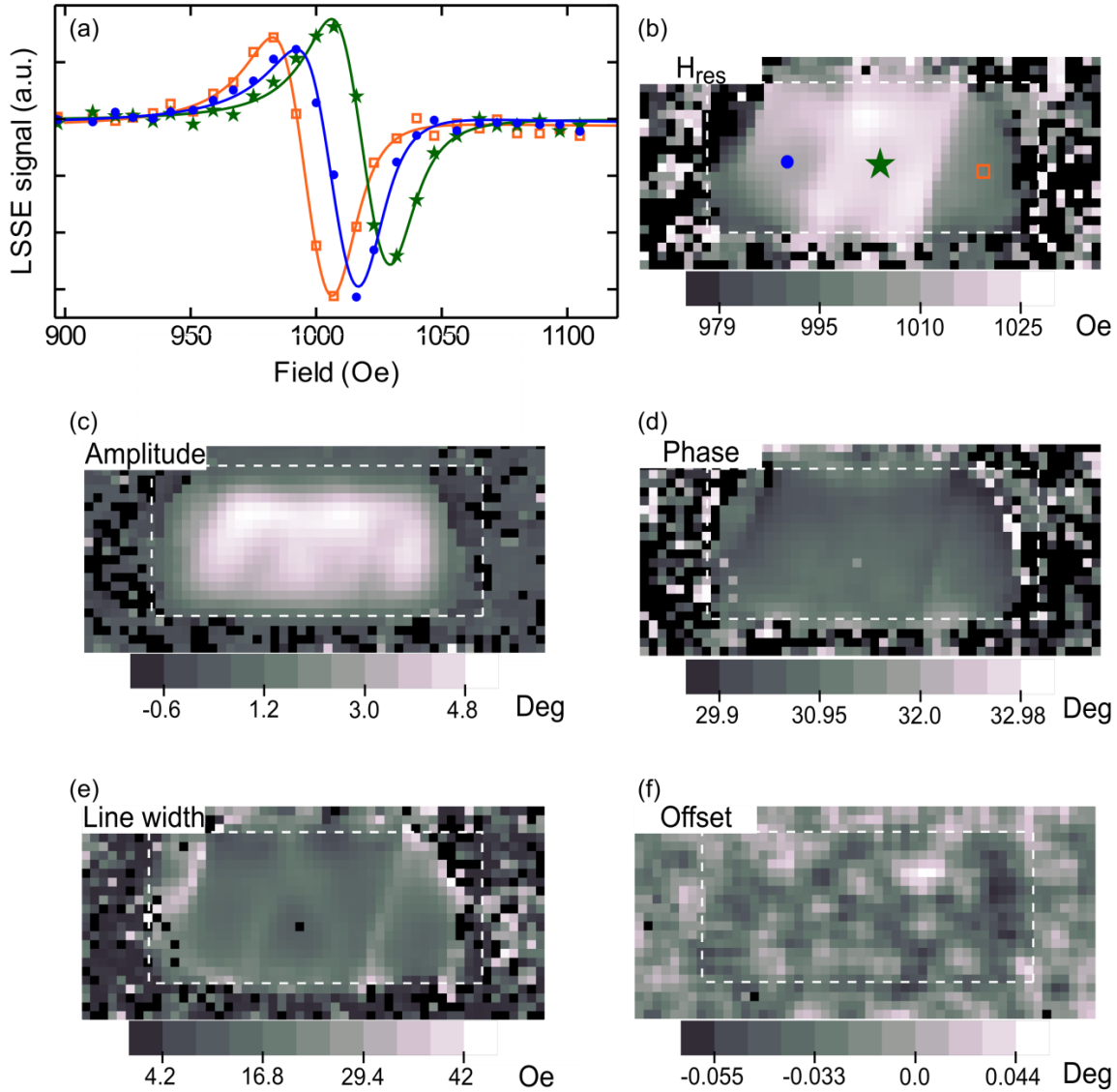

FIG. 6 Spatial maps of FMR fitting parameters for the 4 µm wide wire. (a) Traces are the pixel values of three points on the sample as a function of magnetic field. b-f) Spatial maps of the FMR fitting parameters made by fitting of the FMR curves at each pixel in the sequence of images measured with LSSE. Before fitting, we correct for image-to-image offset and use a 3x3 pixel moving average to smooth the data. (b) Resonance field, the symbols mark the pixels corresponding to the FMR spectra shown in (a). (c) Resonance amplitude, (d) resonance phase, (e) resonance linewidth (f) offset used in the fit.

# Imaging Magnetization Structure and Dynamics in Ultrathin YIG/Pt Bilayers with High Sensitivity Using the Time-Resolved Longitudinal Spin Seebeck Effect

## Supplemental information

Jason M. Bartell[1], Colin L. Jermain[1], Sriharsha V. Aradhya[1], Jack T. Brangham[2], Fengyuan Yang[2], Daniel C. Ralph[1,3], Gregory D. Fuchs[1,3]

[1]Cornell University, Ithaca, NY 14853, USA
[2]Department of Physics, The Ohio State University, Columbus, OH 43016, USA
[3]Kavli Institute at Cornell for Nanoscale Science, Ithaca, NY 14853, USA

**Optical path**

To heat the YIG/Pt bilayers, we use a Ti:Sapphire laser tuned to 780 nm and pulse durations of 3 ps at 76.5 MHz. An electro-optic modulator referenced to the laser pulses is used to reduce the repetition rate to 25.5 MHz, which allows time for thermal recovery. Next, a photoelastic modulator and a polarizer are used to modulate the optical amplitude at 100 kHz for lock-in detection. The resulting vertically polarized light is focused on the sample with a 0.9 NA objective. A fast-steering mirror with a 4-f lens pair is used to scan the laser focus across the sample. The light reflected from the sample is detected with a photodiode bridge.

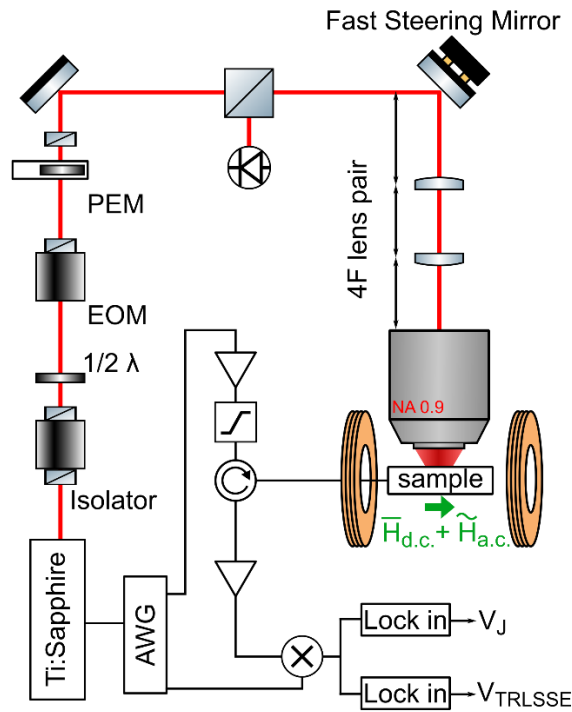

FIG. S1. Schematic of TRLSSE microscope.

**Model of TRLSSE temporal convolution**

We develop a model of the detection circuit to clarify the impact of circuit bandwidth and electrical artifacts on the TRLSSE traces shown in Figs. 2b and 2c. The time domain measurements shown in Fig. 2 show that the duration of $V_{sig}$ matches the ~100 ps duration of the mixing pulse. This implies that thermal gradient induced $V_{LSSE}$ must be sufficiently short-lived to sample the mixing pulse, and thus it is suitable for stroboscopic measurement of GHz frequency dynamics. In addition to the main pulse, we also observe oscillations that can be attributed to non-idealities in the mixing reference pulse produced by the arbitrary waveform generator (AWG) and the RF mixer itself. To account for these effects, we develop a phenomenological model of the signal, which we describe as the convolution of the TRLSSE-induced electrical pulse from the sample and the reference pulse from the AWG as a function of relative delay, $\tau$ [1]. We account for bandwidth contributions and the realistic profile of the mixing reference pulse.

The model consists of a 12 GHz low-pass filter leading to the radio frequency and local oscillator inputs of an idealized mixer (Fig. S2a). The output of the circuit is described by

$$V_{sig}(t) \sim \mathcal{F}^{-1}[\mathcal{F}(V_{mix}) * \mathcal{F}(V_{LSSE}) * LP(f)^2] \tag{S1}$$

Where $LP(f)$ is a first-order low-pass filter $LP(f) = \frac{1}{1+f/f_c}$ for frequency $f$ and cut-off frequency $f_c$ = 12 GHz. The Fourier transform $\mathcal{F}(V)$ is given by $\mathcal{F}(V_{\delta t}) = \frac{1}{\sqrt{T}}\sum_{\delta t=1}^{T} V_{\delta t}\, e^{2\pi i\,(\delta t-1)(f-1)/T}$ where $T$ = 12.9 ns is the duration of the kernel, $\delta t$ = 2.5 ps is the time step, and $f$ is frequency. In the experiment, the mixing pulse $V_{mix}$ is generated by an arbitrary waveform generator (AWG) synchronized to the laser repetition rate with a sampling rate of 9.98 GSamples/s. For mixing voltage pulse $V_{mix}$ we use the output of the AWG measured using a LeCroy SDA 11000 Oscilloscope (Fig. S2b). To model the signal from the sample, $V_{LSSE}$, we use the normalized thermal gradient determined from time-domain finite element (TDFE) calculations (further discussion below). In the main text, we use a 100 ps mixing pulse to acquire the data presented in Fig. 2b,c. and a 600 ps mixing pulse to acquire the other data.

Figure 1 of the main text shows $V_{sig}$ calculated via Eq. S1 normalized to the measured data along with the measured convolution. The model qualitatively captures the oscillations at delay times greater than 100 ps. This model, together with the lack of magnetic field dependence, supports the idea that the oscillations in the data are electrical artifacts, not magnetic oscillations.

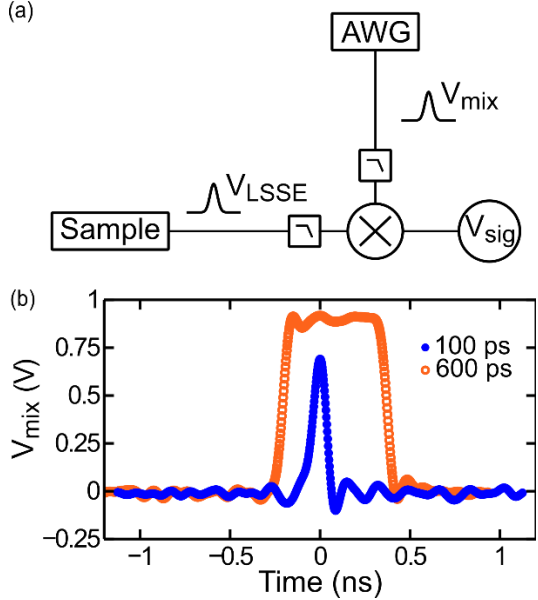

FIG. S2. (a) Schematic of circuit model for interpretation of time-domain circuit. The arbitrary waveform generator (AWG) creates a mixing pulse that goes through a 12 GHz low-pass filter before being mixed with the pulse from the sample that has also been sent through a 12 GHz low-pass filter. (b) Oscilloscope measurements of the mixing pulses used in the experiments.

**Determination of temperature change from laser heating**

Although we know the laser fluence, we do not know the film absorbance for this thin-film limit in which the Pt film is much thinner than the optical skin depth. To determine the temperature change in our experiment we use the following methodology: (1) we numerically calculate the spatiotemporal thermal response to focused laser heating assuming the peak absorbed power is 1 W (an absorbed fluence of 0.7 mJ/cm$^2$). We take the model's predictions for the spatiotemporal thermal evolution to be correct but the total temperature change amplitude as being uncalibrated. (2) We calibrate and measure $V_J$, which is equivalent to using the sample resistivity change as a thermometer. (3) We calculate the $V_J$ from our spatiotemporal thermal model calculations and compare it to the measured $V_J$. We assume there is linear response between the amplitude of the absorbed laser energy and the maximum temperature increase, therefore the ratio of the measured to the calculated values of $V_J$ determines the scale factor of the absorbance. This also scales the temperature increase from the model to a value that agrees with our electrical measurement. Additional details have been described previously in the supporting information of Ref. [1].

We base our model on TDFE calculations of thin-film thermal diffusion to determine the spatiotemporal profile of the thermal gradient temperature distribution. We consider a GGG/YIG(20 nm)/Pt(6 nm) trilayer with material parameters given by Table S1. The YIG/Pt layers are modeled as a 2 µm x 10 µm bar to match the measured device. Heat transfer in the structure is calculated using the diffusion equation

$$\rho\, C_p \frac{\delta T(\mathbf{x},t)}{\delta t} - \kappa \nabla^2 T(\mathbf{x},t) = Q(\mathbf{x},t) \tag{S2}$$

with the COMSOL Multiphysics® software package. In Eq. S2 $\rho$ is the material density, $C_p$, is the specific heat, $\kappa$, is the thermal conductivity, $Q$ is the heat source, $\mathbf{x}$ is the 3D spatial coordinate, and $t$ is time. We assume the YIG/ Pt interfacial thermal conductance is 170 W m$^{-2}$ K$^{-1}$ [2].

We also assume that laser heating only takes place in the Pt layer because of the negligible optical absorption in the YIG [3] and GGG [4]. Thus, the laser is effectively a radially symmetric heat source, with radius $r$, in the platinum with a spatial temporal distribution, for positive z, given by, $Q(\mathbf{x},t) = Exp\left(-\frac{z}{\varepsilon}\right) * \left(\frac{1}{2\pi d^2}\right) * Exp\left(-\frac{r^2}{2 d^2}\right) * Exp\left(-\frac{(t-t_0)^2}{2 w^2}\right)$, where $d = 257$ nm is the focused laser spot size (see "determination of optical spot size" below), $\varepsilon = 12$ nm is the skin depth [5,6], $w = 1.27$ ps is the laser pulse width for a 3 ps FWHM Gaussian pulse, $t_0 = 100$ ps is the time that the heat source is at the maximum. The heat source is applied every 39.6 ns and the simulation runs from time t = 0 ns to t = 42 ns to capture two pulses.

Figure S3 shows the result of the model calculation in the space and time domains. The z-component of the thermal gradient within the YIG decays to 1/e in 92 ps and the temperature difference between the Pt and YIG decays in 91ps, time scales that are experimentally indistinguishable in our measurement and consistent with the time domain measurement shown in Figs 2b,c of the main text. The overall temperature increase within the laser heated region takes longer to relax to room temperature, 295 ps, consistent with Fig. 2d. These calculations support that the TRLSSE signal originates from $\nabla T_z(t)$ (or indistinguishably in this work, the temperature difference between YIG and Pt) and that it is localized in time making it suitable for stroboscopic measurements.

The model calculation predicts about a 400 K change in the Pt, however, as discussed above, we calculated the amplitude of the laser-induced temperature change without experimental knowledge of the absorbed fluence. Therefore, the true temperature change in the Pt may be scaled up or down to account for correct value of the absorbed laser power. To establish the absorbance experimentally, we compare the measured values of $V_J$, which originates from the resistance change of the metal due to laser heating, with a model calculated value of $V_J$, which is determined from the resistance change expected from our model calculation. Specifically, we calculate $V_J$ using the 3D temperature distribution created from laser heating to determine the sample resistance increase. We use the linear relationship between the resistance and the temperature, $R(T) = R_o(1 + \alpha T)$, with the resistance correction factor α = $1.3 \times 10^{-3}$ K$^{-1}$ measured for the Pt films used in our experiment. To compare the calculated value to the experimentally measured $V_J$, we also determine the electrical circuit transfer function in which we account for the measurement bandwidth and gain (see Ref. [1] for further discussion). From this analysis we find that our experimentally measured $V_J$ is 0.12 times the calculated $V_J$, indicating the peak temperature change in the Pt is 50 K, corresponding to a peak absorbed fluence of 0.09 mJ/cm$^2$, 1.6% of the incident laser energy. The uncertainty in the temperature is estimated to be on the order of 25% based on uncertainties in the circuit calibration.

TABLE S1 Material parameters used in the TDFE simulations of laser heating

|     | Specific Heat, $C_p$ (J/kg*K) | Density, $\rho$ (kg/m$^3$) | Thermal conductivity, $\kappa$ (W/m*K) |
| --- | --- | --- | --- |
| Pt  | 133[a] | 21500[a] | 71.6[a] |
| YIG | 570[b] | 5170[c] | 6[b] |
| GGG | 400[b] | 7080[b] | 7.94[b] |

[a]Reference [7]
[b]Reference [8]
[c]Reference [9]

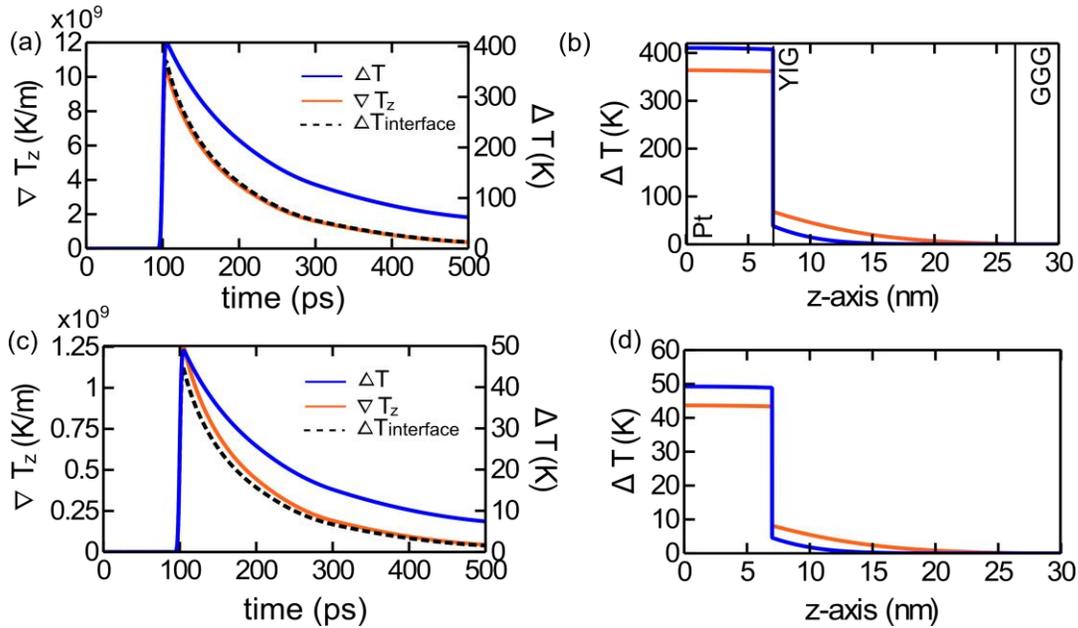

FIG. S3. Time-domain finite element calculations of the temperature and thermal gradient using COMSOL. (a) Time-domain thermal profiles at the YIG/Pt interface calculated with COMSOL assuming an absorbed fluence of 0.7 mJ/cm$^2$ and showing the z-component of thermal gradient in the YIG (orange curve), change in temperature of the Pt (blue curve), and temperature difference between the Pt and the YIG across the interface (black dashed line). The laser turns on at 100 ps in the calculation. (b) Calculated temperature vs. z-axis position showing heating as a function of film depth at the maximum temperature difference (orange curve) and 16 ps later (blue curve). (c,d) The curves from (a) and (b) scaled by the correction factor.

**Effect of interface spin transparency**

The spin Hall magnetoresistance (SMR) is the change in resistance due to spin-dependent transport in a heavy, nonmagnetic metal that shares an interface with a ferromagnet [10]. Thus, for bilayers of the same materials but different spin mixing conductance, measuring SMR provides insight into the efficiency with which spins can cross the interface. The efficiency of interfacial spin transport is important for TRLSSE measurements because in order for the magnetization to be transduced into a voltage, the thermally driven spins must cross the interface.

For the data presented in the main text we find a SMR of 0.063%. We compare the signal from this wire with a relatively strong SMR to the TRLSSE signal from a wire without detectable SMR above the 0.003% noise floor of our lock-in measurement. Both wires were 2 µm x 10 µm with resistances of 296 Ω and 220 Ω for the sample with and without SMR respectivly. The sample without SMR had a thinner YIG film (8 nm), however this is not expected to effect the SMR since SMR is an interfacial effect [11].

Figure S4 shows representative plots of the TRLSSE signal versus field for the different wires at similar laser powers. We find that the sample with SMR has a signal approximately an order of magnitude greater than the sample without. The difference is consistent with the model of TRLSSE driving spin current across the YIG/Pt interface. We also note that even though the signal is reduced, it is still measurable in both samples, enabling measurement of YIG magnetization even in systems that cannot be measured electrically.

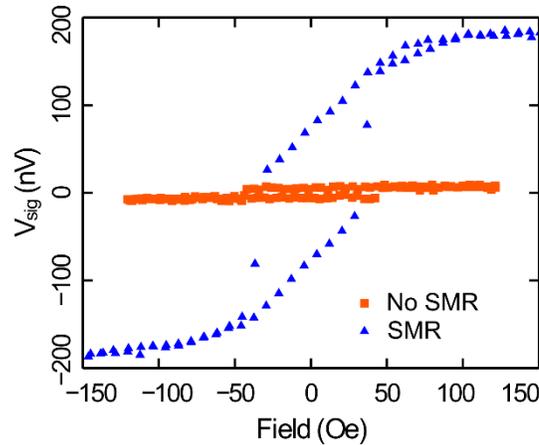

FIG. S4. TRLSSE signal as a function of applied external field for a sample with 0.063% SMR (blue triangles) and a sample with no measurable SMR (orange squares). The applied laser fluences are 5.4 mJ/cm$^2$ and 6.7 mJ/cm$^2$ for the blue an orange curves respectively. For the data presented here, the laser repetition rate was 76.5 MHz and no amplifier was used between the sample and the RF mixer.

### Determination of optical spot size

We determine the diameter of the illuminated area by modeling a Guassian laser focus and fitting the traces of the image shown in Fig. 5a. Fig. S4 shows a y-axis cross section of the image. The trace shows an approximately flat region on the wire surface and a sigmoidal edge due to the convolution of the sharp wire edge with the point-spread function of the laser focus. To fit the reflection signal, $I$, at the edge, we use the convolution of a Guassian with a step function,

$$I = \frac{1}{b\sqrt{2\pi}} \int_{-\infty}^{\infty} \exp\left(-\frac{(x-a)^2}{2b^2}\right) \Theta(x-a) dx, \tag{S3}$$

in which $b$ determines the Guassian width, $a$ is the center of the peak, and $\Theta$ is the step function defined as $\Theta(x-a) = \begin{cases} 0, x < a \\ 1, x \geq a \end{cases}$. The fit of the data yields $b = 0.240 \pm 0.007$ µm and $b = 0.274 \pm 0.010$ µm for the left and right edges respectively. We take the average to be the optical spot size. We attribute the difference between the two edges to a slight out-of-plane tilt of the sample leading to asymmetry in the reflection.

As a comparison, we fit a y-axis scan of the TRLSSE signal to Eq. S3. The result gives $b = 0.380 \pm 0.006$ µm and $b = 0.381 \pm 0.009$ µm for the left and right edges respectively. This difference corresponds to a difference of ~1 pixel between the rise-width of the reflection signal

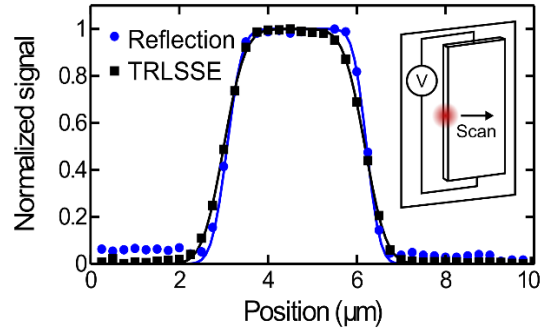

and TRLSSE signal.

FIG. S5. Fit of step edge signal for determination of optical spot size. (a) Line cut in y-axis direction of the reflected light image, shown in Fig. 5a, and the TRLSSE image of the static saturated moment, shown in Fig. 5b. (inset) schematic representation of the sample tilt that can lead to the observed anisotropy.

**Analysis of dynamic TRLSSE images**

To image the ferromagnetic resonance of YIG in the 4 µm wide wire a series of images was taken at fields ranging from 896 to 1105 for an applied RF power of 1.1 mA. A selection of unprocessed images is shown in Fig. 5e-g of the main text. Although the signal is quite clear, we account for sample drift and noise, before fitting the FMR curves.

We correct for sample drift using autocorrelation to find the image overlap. The kernel for the autocorrelation is a 5 ×12.5 µm region from the center of the reflected light image at $H = 896$ Oe (the first image in the series). We determine the drift of subsequent images by finding the distance between the centers of the kernel and the minimum of the autocorrelation. Most of the sample drift is on the order of a pixel (0.25 µm) with a maximum sample drift of $\Delta y = 0.75$ µm and $\Delta x = 0.25$ µm. We correct for the offset by shifting the images and then cropping the borders. The scans cover a large enough area that the cropped region is well away from the wire. After correcting for the sample drift, we remove the background from the vibration edge artifacts by subtracting the TRLSSE signal of the wire at 896 Oe from the subsequent images. Finally, we reduce random pixel to pixel noise, smoothing the signal with a 3x3 pixel moving average. The 3x3 pixel window is approximately the sampling spot size (see determination of optical spot size).

We attribute the small signal features at the edges of the wires in Fig. 5 of the main text to magnetic field modulation induced relative motion between the microscope objective and the sample. As mentioned in the main text, we separate $V_J$ (which is in principle non-magnetic) from $V_{TRLSSE}$ (which is magnetic) by adding a modulation magnetic field (7.6 Oe RMS, $\omega_H = 383$ Hz) to the d.c. magnetic field. We then demodulate $V_{sig}$ with respect to $\omega_H$ using a lock-in amplifier. Although this procedure is effective for isolating $V_{TRLSSE}$ from $V_J$ when we focus in the center of the wire (away from the wire edge), the modulation field induces a tiny "wobble" in the laser focus on the sample. When the laser is focused on the sample edge and a current is applied to the sample, the wobble introduces a slight modulation of $V_J$ at $\omega_H$ because $\frac{dV_J}{dH} = (\frac{dV_J}{dy})(\frac{dy}{dH})$, where $\frac{dy}{dH}$ is due to field-induced mechanical motion and $\frac{dV_J}{dy}$ is large at the sample edge. We note that these edge signals are independent of external field but that they are sensitive to the current amplitude and phase, both of which are consistent with this interpretation of the artifact. In Fig. S6 we plot both the profile of the externally modulated field signal in the y-direction and the numerical derivative of $V_J$ measured by the lock-in referenced to the 100 kHz laser modulation rate, which demonstrates their correspondence.

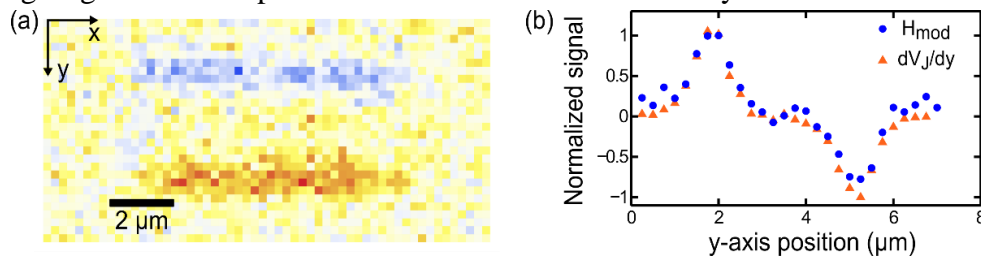

FIG. S6. (a) Spatial variation of the TRLSSE in a 4 × 10 μm YIG/Pt wire at 911 Oe. The signal measured by a lock-in amplifier referenced to the frequency of an a.c. magnetic field. (b) Profile of the TRLSSE signal shown in (a) (blue circles) and the derivative of $V_J$ from the same area of the wire (orange triangles). The trace is the average of twenty-six y-axis line scans from along the length of the wire.